\documentclass[usenatbib]{mn2e}
\usepackage{graphicx}
\usepackage{epstopdf}
\usepackage{natbib}
\usepackage{verbatim}
\usepackage{amssymb,amsmath}
\usepackage{ulem}
\usepackage[bookmarks,bookmarksnumbered,colorlinks=true, citecolor=blue, linkcolor=black, draft=false]{hyperref}

 \usepackage{times}

\usepackage{color}
\usepackage[mathscr]{eucal}

\newcommand\apj{ApJ}
\newcommand\apjl{ApJ}
\newcommand\aap{A\&A}
\newcommand\mnras{MNRAS}
\newcommand\pasj{PASJ}
\newcommand\nat{Nature}

\title[SFR and metallicity time variability drive the FMR]{Similar star formation rate and metallicity evolution timescales drive the fundamental metallicity relation}

\author[P. Torrey et al.]
       {\parbox{18cm}{Paul~Torrey$^{1}$\thanks{ptorrey@mit.edu}\thanks{Hubble Fellow}, 
       Mark~Vogelsberger$^{1}$\thanks{Alfred P. Sloan Fellow},
       Lars~Hernquist$^{2}$,
       Ryan~McKinnon$^{1}$,
       Federico Marinacci$^{1}$,
       Robert~A.~Simcoe$^{1}$,
       Volker Springel$^{3,4,5}$,
       Annalisa Pillepich$^{6}$,
       Jill Naiman$^{2}$, \\
       R\"udiger Pakmor$^{3}$,
       Rainer Weinberger$^{3}$,
       Dylan Nelson$^{5}$,
       Shy Genel$^{7,8}$
       }\vspace{0.3cm}\\ 
         $^1${MIT Kavli Institute for Astrophysics \& Space Research, Cambridge, MA, 02139, USA}\\
         $^2${Harvard--Smithsonian Center for Astrophysics, 60 Garden Street, Cambridge, MA 02138}\\
         $^3${Heidelberg Institute for Theoretical Studies, Schloss-Wolfsbrunnenweg 35, D-69118 Heidelberg, Germany}\\
         $^4${Zentrum f{\"u}r Astronomie der Universit{\"a}t Heidelberg, ARI, M{\"o}nchhofstr. 12-14, D-69120 Heidelberg, Germany}\\
	$^5${Max-Planck-Institut f{\"u}r Astrophysik, Karl-Schwarzschild-Str. 1, 85741 Garching, Germany}\\
	$^6${Max-Planck-Institut f{\"u}r Astronomie, K{\"o}nigstuhl 17, 69117 Heidelberg, Germany}\\
	$^{7}$Center for Computational Astrophysics, Flatiron Institute, 162 Fifth Avenue, New York, NY 10010, USA\\
	$^{8}$Columbia Astrophysics Laboratory, Columbia University, 550 West 120th Street, New York, NY 10027, USA\\
         }

\begin{document}

\maketitle

\vskip 100mm

\begin{abstract}
The fundamental metallicity relation (FMR) is a postulated correlation between galaxy stellar mass, star formation rate (SFR), and gas-phase metallicity.
At its core, this relation posits that offsets from the mass-metallicity relation (MZR) at a fixed stellar mass are correlated with galactic SFR.
In this Letter, we quantify the timescale with which galactic SFRs and metallicities evolve using hydrodynamical simulations.
We find that Illustris and IllustrisTNG predict that galaxy offsets from the star formation main sequence and MZR evolve over similar timescales,
are often anti-correlated in their evolution,
evolve with the halo dynamical time, 
and produce a pronounced FMR.
In fact, for a FMR to exist, the metallicity and SFR must evolve in an anti-correlated sense which requires that they evolve with similar time variability.
In contrast to Illustris and IllustrisTNG, we speculate that the SFR and metallicity evolution tracks may become decoupled in galaxy formation models dominated by globally-bursty SFR histories, which could weaken the FMR residual correlation strength.
This opens the possibility of discriminating between bursty and non-bursty feedback models based on the strength and persistence of the FMR -- especially at high redshift. 
\end{abstract}


\renewcommand{\thefootnote}{\fnsymbol{footnote}}
 
 \begin{figure*}
\centerline{\vbox{\hbox{ 
\includegraphics[width=0.45\textwidth]{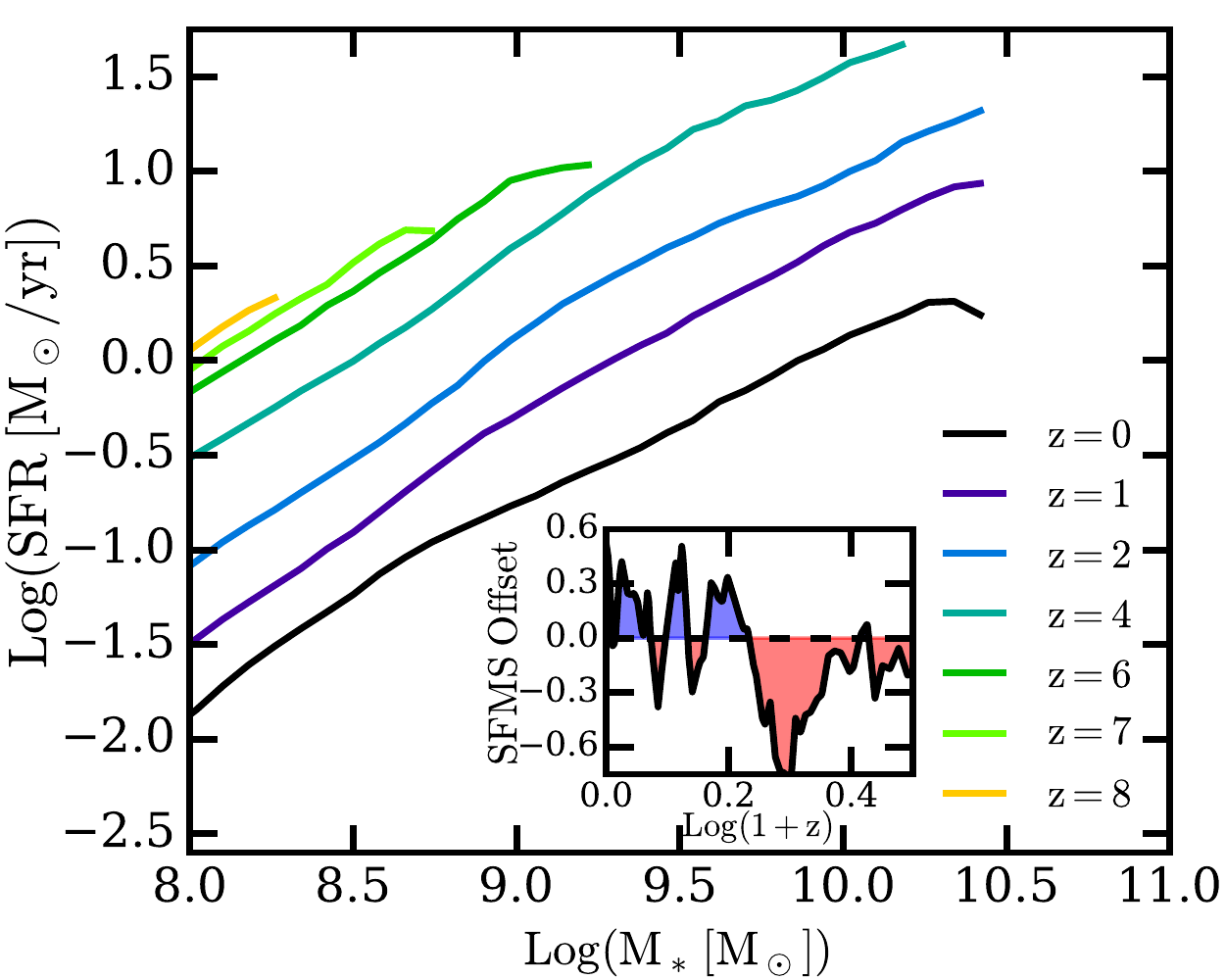}
\includegraphics[width=0.45\textwidth]{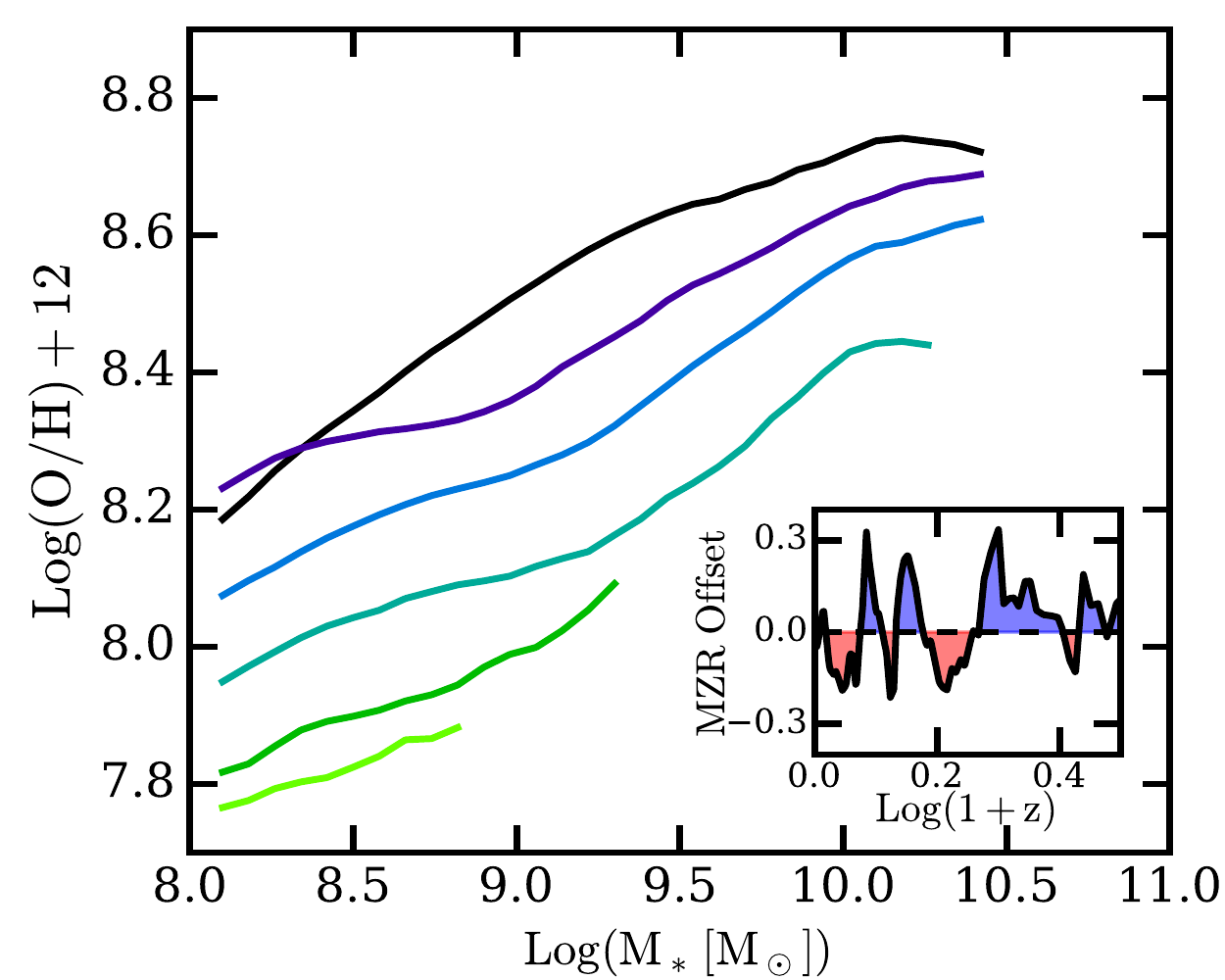}
 }}}
\caption{The median SFMS (left) and MZR (right) for the TNG100 simulation at several redshifts.
The average SFR (metallicity) evolves toward higher (lower) values with increasing redshift at a fixed mass scale.
Inset panels show evolution tracks of offset from the SFMS (left) and MZR (right) for one example galaxy.
Blue/red regions highlight when the galaxy is above/below the mean relation.
 }
\label{fig:median_relations}
\end{figure*}

\section{Introduction}
The mass metallicity relationship (MZR) is an important galaxy scaling relation that describes the coevolution of galaxies and their metal content~\citep[e.g.][]{Tremonti2004}.
Over the past several years, observational results have begun to indicate that the scatter in the MZR may be correlated with galactic star formation rates~\citep[SFR; e.g.][]{Ellison_FMR} or gas masses~\citep[e.g.][]{Bothwell2013}. 
This has led to the suggestion that there is a fundamental metallicity relation~\citep[FMR;][]{LaraLopez2010, Mannucci_FMR} describing the combined relation between galactic stellar mass, SFR (or gas mass), and metallicity. 
Evidence has been presented showing that this FMR holds across a wide mass range~\citep{Salim2014, Brown2017} as well as out to high redshift~\citep{Belli2013, Stott2014,  Bothwell2016a} with increasingly systematic and comprehensive analyses~\citep{Sanders2017}.
Further, theoretical models have been built explaining why the FMR should naturally occur~\citep[e.g.][]{Lilly2013, Forbes2014, Zahid2014}.

However, there is no consensus about the existence of the FMR~\citep[e.g.][]{Sanchez2013, Sanchez2017}.  
In particular, there are concerns that the FMR is driven by systematic uncertainties in nebular emission line metallicity diagnostics~\citep{Telford2016} 
or contaminated by incomplete/non-global fibre corrections~\citep{BarreraBallesteros2017, Ellison2017, Sanchez2017}.
There is also disagreement about the strength~\citep[e.g.][]{Andrews2013} or mass dependence~\citep[e.g.][]{Yates2012} of the residual correlation as well as the persistence of the FMR in high redshift data~\citep{Yabe2015}.
In short, while the FMR provides an inviting link between the fluctuations in the metallicities and SFRs (or gas masses), it is not yet clear if the observational evidence fully supports this scenario.

Yet, in a recent paper~\citep{Torrey2017}, we demonstrated that a FMR is naturally produced in hydrodynamical simulations~\citep[see also][]{Dave2017, DeRossi2017}.
In this Letter, we explore the conditions that are required for a FMR to emerge.
In particular, the emergence of this relation relies on galactic \textit{offsets} from the MZR and star formation main sequence (SFMS) remaining anti-correlated as they evolve, which we argue requires that the SFR and metallicity share similar dominant evolution timescales.
If either the SFR or metallicity evolved much faster than its counterpart, then the anti-correlation between the two quantities would be washed out.

As we argue in this Letter, the dominant timescales for metallicity and SFR evolution are similar in our galaxy formation model.
We further argue that the evolution timescales for SFRs and metallicities are generally set by galaxy dynamics in our model, not by the adopted feedback physics.
The similarity in the SFR and metallicity evolution timescales enables the existence of the FMR.
However, as we briefly address in Section~\ref{sec:Conclusions}, we speculate that the strength of the FMR -- especially for low mass galaxies at high redshift -- may be reduced in models with particularly strong and/or globally-bursty feedback.

\vspace{-5 mm}

\section{Methods}
\label{sec:Methods}
In this Letter we analyze the time variability of galactic SFRs and metallicities using the IllustrisTNG simulation suite~\citep{Marinacci2017, Naiman2017, Nelson2017, Pillepich2017b, Springel2017}.
The IllustrisTNG simulation suite builds on the original Illustris simulation via a series of numerical and physical model improvements~\citep{Weinberger2017, Pillepich2017} over the original Illustris model~\citep{Vogelsberger2013, Torrey2014}.
In this Letter we use the TNG100 simulation which employs a simulation box of side length $L\approx100\; \mathrm{Mpc}$ and is an analog to the original Illustris simulation volume.

SFRs are determined using a subgrid model which leads to smoothly-varying, non-bursty SFR histories~\citep{SH03, vogelsberger2014a, genel2014, Sparre2015}.
Gas is converted into stars using a stochastic star formation prescription.
Each simulation stellar particle represents an unresolved full stellar population which we assume is described by a Chabrier initial mass function.
As stellar particles age, they return both mass and metals locally to the ISM resulting in time- and spatially-dependent metal enrichment.
Enrichment predictions of the IllustrisTNG simulations agree on galactic~\citep{Naiman2017, Torrey2017} and cluster scales~\citep{Vogelsberger2017} with observations.
Variability in the SFRs and metallicity values is therefore naturally driven by the variety of formation histories among the simulated galaxy population.
We always quote instantaneous (un-smoothed) SFRs and metallicity values as the SFR weighted average metallicity for all gas within a galaxy.

\begin{figure}
\centerline{\vbox{\hbox{ 
\includegraphics[width=0.45\textwidth]{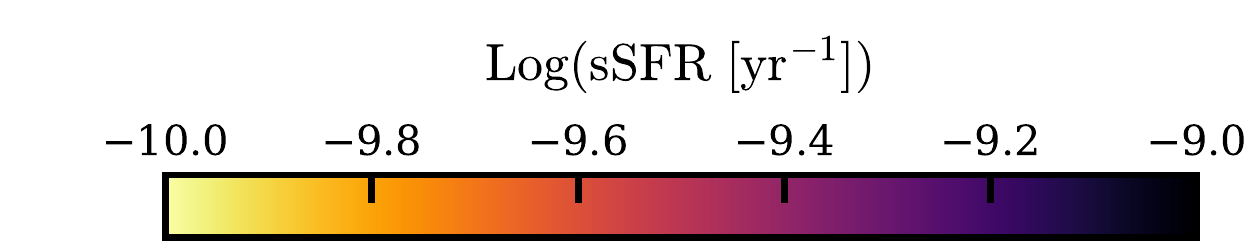} }}}
 \centerline{\vbox{\hbox{ 
\includegraphics[width=0.45\textwidth]{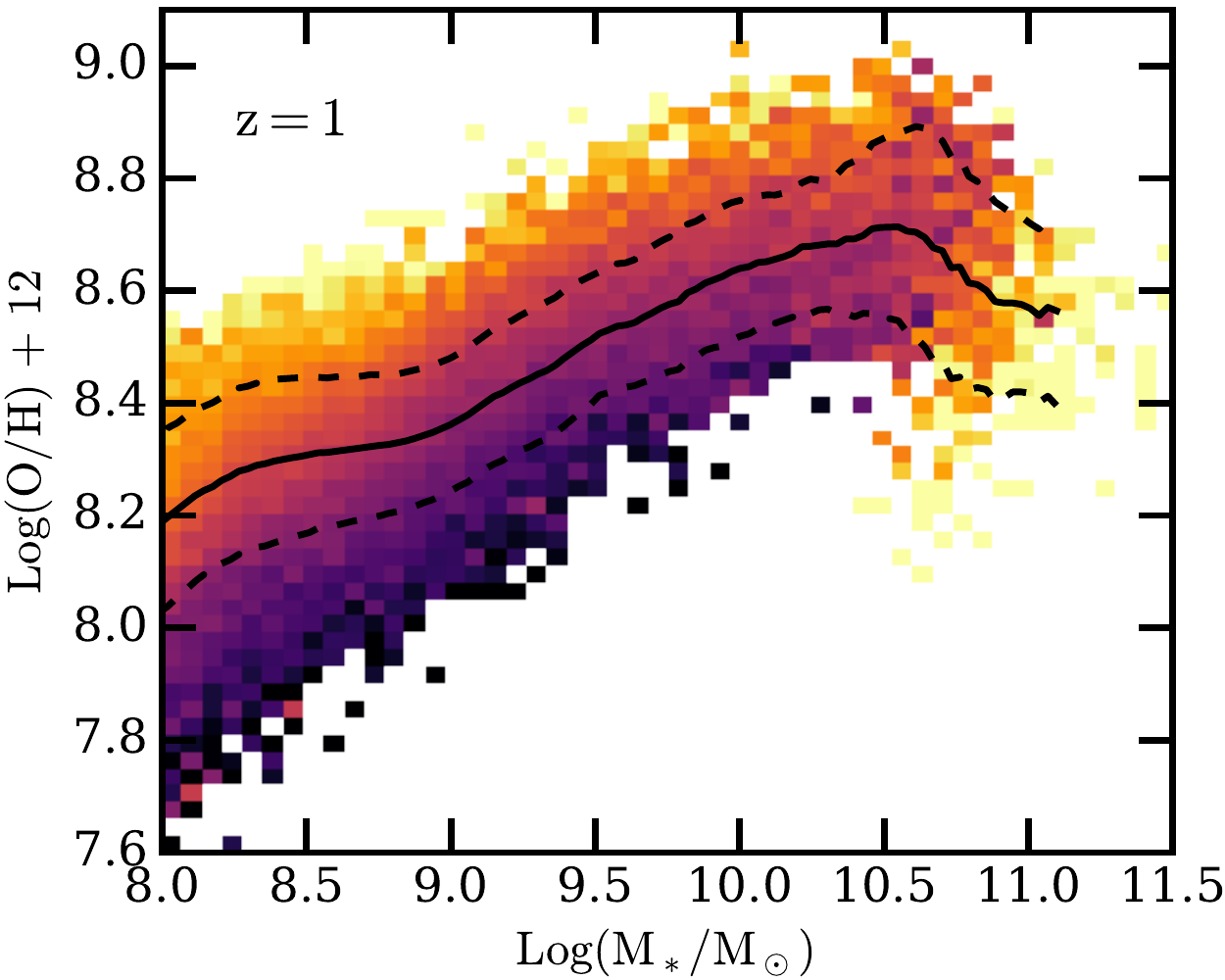} }}}
\caption{ 
Two-dimensional histogram of average specific SFRs for IllustrisTNG galaxies as a function of metallicity and stellar mass at $z=1$.
Solid and dashed black lines indicate the median MZR and one sigma scatter, respectively.
There is a residual correlation about the MZR where galaxies with high metallicities have low SFRs, and galaxies with low metallicities have high SFRs.
 }
\label{fig:FMR}
\end{figure}

\vspace{-5 mm}

\section{Results}
\label{sec:Results}
Figure~\ref{fig:median_relations} shows the SFMS (left panel) and MZR (right panel) at several redshifts.
There is significant redshift evolution in both of these relations with higher redshift galaxies having higher SFRs~\citep{Weinberger2017b} and lower metallicities~\citep{Torrey2017}.
We use these relations to build a two-dimensional interpolation function that yields the SFR or metallicity as a function of stellar mass and redshift.

Using the median relations defined above, we track galaxies in time and identify their offsets from the scaling relations.
The insets in Figure~\ref{fig:median_relations} show tracks for the offset evolution of one randomly selected galaxy with mass $M_* \approx 10^{10} \mathrm{M}_\odot$ at $z=0$.
We highlight periods of time where this galaxy is above or below the scaling relations by shading blue or red, respectively.
There are some features in the offset evolution tracks that are repeated between the SFR and metallicity panels.  
In particular, we find ``mirrored"  offset evolution between the metallicity and SFR offset values 
over a broad redshift range where enhancements in the SFR of this galaxy above the SFMS correspond to depressions in the metallicity below the MZR.

The anti-correlated evolution of the MZR and SFMS offsets has the observable consequence of driving residual correlations in the scatter about the MZR.
Figure~\ref{fig:FMR} shows the average specific SFR for galaxies distributed about the simulated MZR at redshift $z=1$.
There is a clear residual correlation between MZR offset and SFR for the full galaxy population, indicating that the anti-correlated offset behavior shown in the individual evolution tracks in the inset of Figure~\ref{fig:median_relations} is commonly found in the larger galaxy population.
\textcolor{black}{Although we only plot $z=1$ here for brevity, this residual trend is also present at other redshifts~\citep{Torrey2017} and we summarize the best fitting slope relating offset from the MZR with offset from the SFMS in Table~\ref{table:FMR_slopes}.
If galaxies randomly changed their SFR and metallicity values, or if the SFR and metallicity evolution had significantly different time variability, the strength of the residual correlation between SFR and MZR offset would be washed out.}

In this Letter we use a simple metric for identifying the smallest timescale with significant SFR variability by determining how well a galaxy population's offsets from the SFMS at some time, $t_0$, indicate the same galaxy population's offsets from the SFMS at some earlier time, $t_0 - \Delta t$.
We specifically use a Pearson correlation coefficient, $\rho(M_*, t_0, \Delta t)$, to describe the strength of correlation between the offsets from the SFMS for a galaxy population with average stellar mass, $M_*$, measured at the two different times.
The correlation strength is measured as a function of initial time, $t_0$, time separation, $\Delta t$, and in a series of initial stellar mass bins $M_*\pm\Delta 0.25$ dex.
For any given initial time, $t_0$, and initial stellar mass, $M_*$, we fit exponential decay curves, $\rho(M_*, t_0, \Delta t) = \mathrm{exp}(-\Delta t/\tau)$, to the measured correlation coefficients as a function of time separation, which we find provides an adequate fit.
We use the best fitting decay timescale, $\tau=\tau(M_*, t_0)$, to define the SFR evolution timescale for that particular initial time, $t_0$, and initial stellar mass, $M_*$.
We perform an identical procedure to determine metallicity evolution timescales, where we instead use offsets from the MZR as input.
There are $\sim$80 simulation snapshots over the redshift range $0\leq z \leq4$ that are used in this procedure.

\begin{table}
\begin{center}
\caption{Best fitting slopes, $\alpha$, between SFMS offset, $\Delta \mathrm{log\;SFR} $, and MZR offset, $\Delta \mathrm{log}\;Z$, in $\pm0.1$ dex stellar mass bins at several redshifts.  Slopes are calculated as $\Delta \mathrm{log}\;Z = \alpha \Delta \mathrm{log\;SFR} $  via $\chi^2$ minimization. }
\label{table:FMR_slopes}
\begin{tabular}{ l c c c c  }
\hline
$M_*$		  			& $ 10^9 \mathrm{M}_\odot$		& $ 10^{9.5} \mathrm{M}_\odot$	& $10^{10} \mathrm{M}_\odot$	&  $ 10^{10.5} \mathrm{M}_\odot$	 	\\ 
\hline
\hline
$\alpha(z=0)$			& -0.19  						& -0.30 						& -0.30					& -0.11							\\
$\alpha(z=1)$			& -0.28  						& -0.28 						& -0.27					& -0.24							\\
$\alpha(z=2)$			& -0.27  						& -0.29 						& -0.25					& -0.29							\\
$\alpha(z=3)$			& -0.30  						& -0.27 						& -0.34					& --								\\
$\alpha(z=4)$			& -0.26  						& -0.26 						& --						& --								\\
\hline
\hline
\end{tabular}
\end{center}
\end{table}

We note that this simple method only identifies the shortest timescale over which there is significant variation in the SFR values.
Noisy data may lead to a short derived evolution timescale, even if there is meaningful/significant longer term variability.
However, for the smoothly varying SFR and metallicity evolution histories typical of IllustrisTNG galaxies, we find this method adequately describes the physically relevant evolution/variability timescales.

\begin{figure*}
\centerline{\vbox{\hbox{
\includegraphics[width=0.45\textwidth]{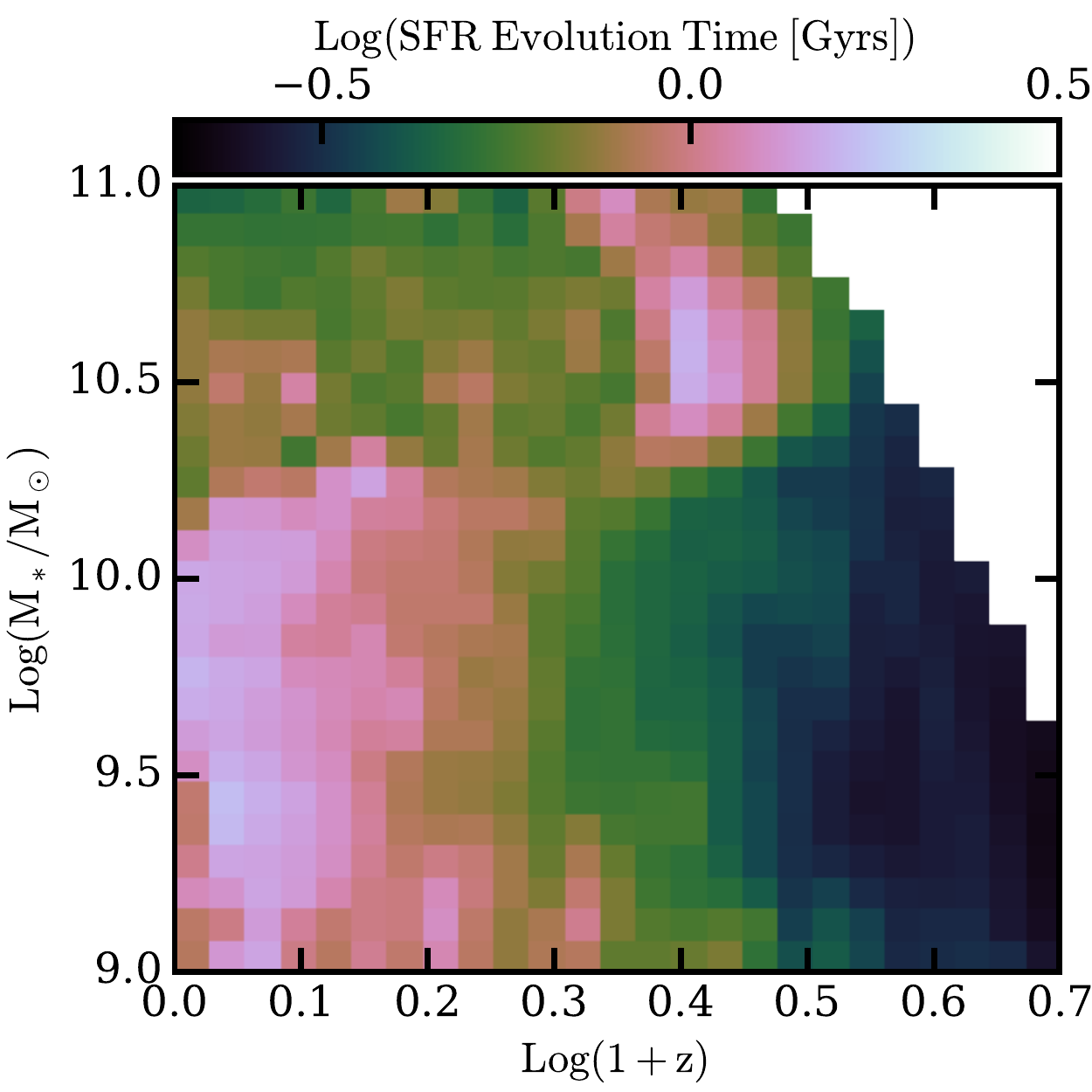}    
\includegraphics[width=0.45\textwidth]{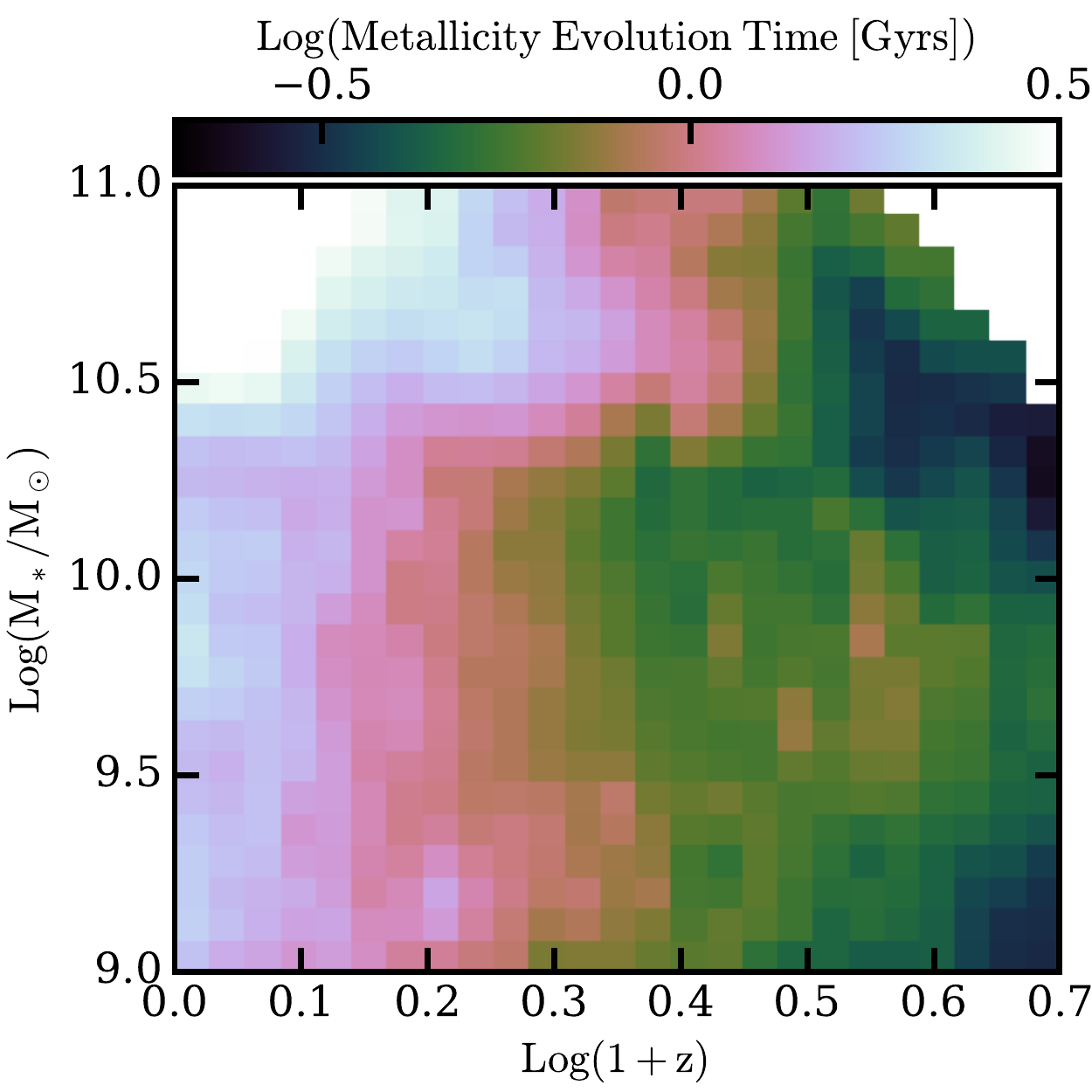}    
}}}
\caption{
Maps of the evolution timescale (see text for details) for the SFRs (left) and metallicities (right) in the TNG100 simulation.
In general, the SFR correlation timescales are somewhat shorter than the metallicity timescales, but both show a redshift dependence where higher redshift galaxies evolve on shorter timescales for galaxies with $M_* \lesssim 10^{10.5} \mathrm{M}_\odot$.  }
\label{fig:correlation_time_map}
\end{figure*}

Figure~\ref{fig:correlation_time_map} shows the derived SFR (left) and metallicity (right) evolution timescales as a function of stellar mass and redshift for the full galaxy population.
We highlight two trends identifiable within this space.
First, the SFR and metallicity timescales associated with the highest mass galaxies (i.e. $M_* > 10^{10.5} \mathrm{M}_\odot$) are somewhat disjoint from the rest of the galaxy population owing to active galactic nuclei (AGN) feedback.
In this regime, SFRs can change rapidly as AGN feedback operates, but metallicities evolve more slowly as the low SFRs and low accretion rates make it difficult to modify the metallicity of the central gas reservoir.  
Second, the timescales that describe the SFR and metallicity evolution for galaxies with stellar masses below $M_* \lesssim 10^{10.5} \mathrm{M}_\odot$ are similar.
There is a trend where higher redshift galaxies have shorter SFR and metallicity evolution timescales compared to their lower redshift companions.
While $z=0$ galaxies have evolution timescales of just over a Gyr, high redshift galaxies of the same mass have evolution timescales several times shorter than this.
There is limited change in the evolution timescale with mass at a fixed redshift for both the metallicity and SFR maps for masses $M_* \lesssim 10^{10.5} \mathrm{M}_\odot$.

\begin{figure}
\centerline{\vbox{\hbox{
\includegraphics[width=0.5\textwidth]{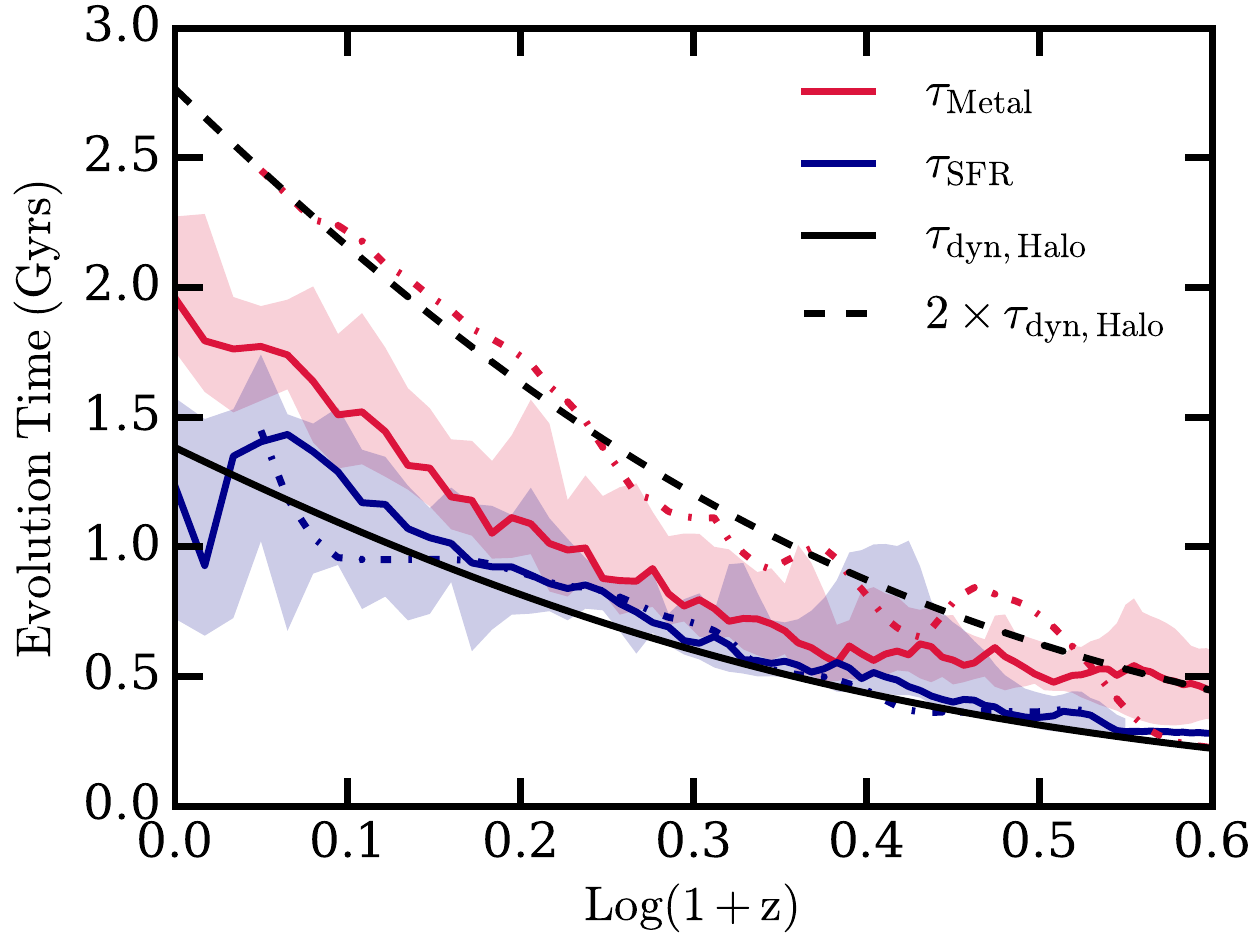}    
}}}
\caption{ Metallicity (red) and SFR (blue) evolution timescales \textcolor{black}{ averaged over the mass range $10^9 < M_*/\mathrm{M}_\odot < 10^{10.5}$ } as a function of redshift for IllustrisTNG (solid) and Illustris (dot-dashed).    Additionally, the halo dynamical time (black solid line) and twice the halo dynamical time (black dashed) lines are shown.
 }
\label{fig:correlation_time_series}
\end{figure}

Figure~\ref{fig:correlation_time_series} shows a direct comparison of the redshift evolution of the SFR and metallicity correlation timescales.
The colored solid lines and shaded regions indicate stacked median and 10${}^{\mathrm{th}}$/90${}^{\mathrm{th}}$ percentile ranges for the evolution timescale evenly weighted over the mass range $10^9 \mathrm{M}_\odot < M_* < 10^{10.5} \mathrm{M}_\odot$.
The scatter about these evolutionary tracks is remarkably small considering the large mass range, but is consistent with the limited mass dependence shown in both panels of Figure~\ref{fig:correlation_time_map}.
Interestingly, we find that the metallicity and SFR evolution timescales are similar in magnitude and evolve in a similar sense with time.
The metallicity evolution timescales are somewhat larger (i.e. $\sim1.2-1.5$ times larger) than the SFR evolution timescales.
There is a late time drop in the SFR correlation timescale which is dominated by the highest mass bins (see Figure~\ref{fig:correlation_time_map}) where AGN feedback plays an increasing role in modulating the SFRs.

The blue and red dot-dashed lines in Figure~\ref{fig:correlation_time_series} indicate the Illustris SFR and metallicity evolution timescales.
While the Illustris SFR evolution timescales are a very good match to the IllustrisTNG trends, the Illustris metallicity evolution timescales are roughly $\sim$30 per cent larger than the IllustrisTNG metallicity evolution values.
Overall, however, the timescales of both models show a similar trend in their redshift evolution and all measured SFR and metallicity  evolution timescales are the same within a factor of $\sim2$.
Although not shown here, we note that Illustris also shows a Z-SFR anti-correlation~\citep{Genel2016}.

In addition to the metallicity and SFR evolution timescales, Figure~\ref{fig:correlation_time_series} shows the halo dynamical time
\begin{equation}
\tau_{\mathrm{DM,dyn}} = \left( \frac{3 \pi}{32 G \rho_{200,\mathrm{crit} } } \right)^{1/2} \sim 0.1 \; \tau_{\mathrm{H}},
\end{equation}
where $\tau_{\mathrm{H}}$ is the Hubble time.
Without adjustment, the halo dynamical time tracks the SFR evolution timescales for both simulations reasonably well.
The metallicity evolution timescales for IllustrisTNG and Illustris also follow the halo dynamical time tracks, but are scaled up by factors of 1.5 and 2.0, respectively.
The scaling of these quantities with the halo dynamical time points to a picture where variability in the SFRs and metallicities is dominated by the evolution of the halo~\citep{Lilly2013} -- not by the specific adopted Illustris or IllustrisTNG feedback physics.
\textcolor{black}{This conclusion is supported because the halo dynamical time has no mass dependence and the measured SFR/metallicity evolution timescales have only a limited/modest mass dependence.}
The variation in the metallicity evolution timescales between Illustris and IllustrisTNG does, of course, indicate that the implemented galaxy formation physics can impact the evolution timescales at the factor of $\sim2$ level.
\textcolor{black}{However, while the FMR is not significantly impacted by factor of $\sim$2 differences in the SFR and metallicity evolution timescales~\citep{Genel2016, Torrey2017},
as we briefly discuss in the following section, we speculate that models with order-of-magnitude different/shorter evolution timescales may have a weaker FMR since 
very rapid SFR variability will wash out residual correlations between SFR and metallicity offsets.}

\vspace{-8 mm}

\section{Discussion and Conclusions}
\label{sec:Conclusions}

In this Letter, we have shown that SFRs and metallicities in the Illustris and IllustrisTNG models evolve over similar timescales which are reasonably well matched to the halo dynamical time.
This similarity allows for the existence of the FMR.
The mirrored nature of the offsets from the MZR and SFMS that drive the correlation between offset from the MZR and SFR only exists because the metallicity and SFRs of galaxies evolve over \textit{similar} timescales.
If the dominant timescales for SFR and metallicity evolution were very different, then residual correlations with SFR about the MZR would be weakened.
The existence of correlated scatter as described by the FMR instead implies that the dominant timescale for variation with respect to the mean relations must be similar for galactic metallicity and SFRs.

Our models predict a continued existence of the FMR out to high redshift~\citep{Torrey2017} because the employed model gives rise to non-bursty SFR histories that evolve on timescales comparable to the halo dynamical time~\citep[see also, e.g.,][]{Dave2017, DeRossi2017}.
We do not necessarily expect that the results presented in this Letter would be recovered by simulations with galaxies dominated by globally-bursty SF histories.
For example, globally-bursty stellar feedback can strongly impact SFRs, driving significant changes to the SFR evolution without necessarily impacting metallicity in the same way.
Specifically, outflows possessing the same metallicity as the ISM can immediately change the SFR of a galaxy, while not impacting the ISM metallicity.
Globally-bursty SF histories have the ability to drive SFR evolution timescales down -- possibly by orders of magnitude to $\sim$10 or 100 Myrs~\citep[e.g.][Figure 9]{Sparre2017} -- which could drive the strength of correlation between offset from the MZR and SFR to be weaker than what our models find.

As of yet, the existence or strength of the FMR is still under debate -- both at low and high redshift -- and so it is not clear which of these models is a better match to the real Universe.
Accurately assessing the existence or strength of the FMR, especially toward higher redshift, is important because it may discriminate between bursty and non-bursty galaxy formation feedback models.

\vspace{-6 mm}

 \section*{ACKNOWLEDGEMENTS}
 
PT thanks Alice Shapley, Ryan Sanders, and Tiantian Yuan for helpful discussion.
PT acknowledges support from NASA through Hubble Fellowship grant HST-HF2-51341.001-A awarded by STScI, which is operated under contract NAS5-26555. 
RM acknowledges support from the DOE CSGF under grant number DE-FG02-97ER25308.
RW, VS and RP acknowledge support through the ERC under ERCStG grant EXAGAL-308037, and the Klaus Tschira Foundation. 
RW acknowledges the IMPRS for Astronomy and Cosmic Physics at the University of Heidelberg. 
VS acknowledges support from subproject EXAMAG of the Priority Programme 1648 \textit{Software for Exascale Computing} of the German Science Foundation. 
MV acknowledges the support of the Alfred P. Sloan Foundation and NASA ATP grant NNX17AG29G. 
JPN acknowledges support of NSF AARF award AST-1402480. 
The Flatiron Institute is supported by the Simons Foundation. 
The IllustrisTNG simulations and exploratory runs were run on the HazelHen Cray XC40-system (project GCS-ILLU),  Stampede at TACC/XSEDE (allocation AST140063), and Hydra and Draco at the Max Planck Computing and Data Facility.

\vspace{-6 mm}


\end{document}